\documentclass[preprint,5p,times,twocolumn,authoryear]{elsarticle}
\usepackage{amsmath,amssymb}
\usepackage{amsthm}
\usepackage[all]{xy}

\usepackage{pstricks}
\usepackage{pst-plot}
\usepackage{booktabs,threeparttable}

\usepackage{array}

\newcolumntype{x}[1]{>{\centering\hspace{0pt}}p{#1}}

\newdefinition{definition}{Definition}

\newproof{pf}{Proof}

\begin{document}

\begin{frontmatter}

\title{Characterizing Assumption of Rationality by Incomplete Information}

\author{Shuige Liu}
\ead{shuige\_liu@aoni.waseda.jp}
\address{Faculty of Political Science and Economics, Waseda University, Japan}

\begin{abstract}
We characterize common assumption of rationality of 2-person games within an
incomplete information framework. We use the lexicographic model with
incomplete information and show that a belief hierarchy expresses common
assumption of rationality within a complete information framework if and
only if there is a belief hierarchy within the corresponding incomplete
information framework that expresses common full belief in caution,
rationality, every good choice is supported, and prior belief in the
original utility functions. 
\end{abstract}


\end{frontmatter}

\section{Introduction\label{sec:int}}
Assumption of rationality is a concept in epistemic game theory introduced
by Brandenburger et al. \cite{bfk08} and studied in Perea \cite{p12} by
using lexicographic belief. A lexicographic belief is said to \emph{assume
the opponents' rationality} means that a \textquotedblleft
good\textquotedblright\ choice always occurs in front of a \textquotedblleft
bad\textquotedblright\ one. Here by good we mean a choice of the opponent
can be supported by a \emph{cautious} belief of him, that is, a belief that
does not exclude any choice of the opponents; by bad we mean it cannot be
supported by any such belief.

Like other concepts in epistemic game theory such as permissibility
(Brandenburger \cite{b92}) and proper rationalizability (Schuhmacher \cite%
{sc99}, Ascheim \cite{a01}), iterative admissibility is defined partly to
alleviate the tension between caution and rationality (Blume et al. \cite%
{bbd91a}, Brandenburger \cite{b92}, B\"{o}rgers \cite{bo94}, Samuelson \cite%
{s92}, B\"{o}rgers and Samuelson \cite{bs94}) by sacrificing rationality.
Caution requires that every choice, be it rational or not, should appear in
a belief; assumption of rationality only requires that those rational
choices should occur in front of those irrational ones but cannot exclude
the irrational ones. On the other hand, since rationality is a basic
assumption on human behavior in game theory, it seems desirable to find an
approach to have a \textquotedblleft complete\textquotedblright\ rationality
while to keep the definition of iterative admissibility.

One approach is to use an incomplete information framework introduced by
Perea and Roy \cite{ps17} which defined standard probabilistic epistemic
model with incomplete information and used it to characterized $\varepsilon $%
-proper rationalizability. Following their approach, Liu \cite{l17} defined
lexicographic epistemic models with incomplete information, constructed a
mapping between them and models with complete information, and characterized
permissibility and proper rationalizability. In this paper, we still use the
construction in Liu \cite{l17} and characterize assumption of rationality.
We show that a choice is optimal for a belief hierarchy which expresses
common assumption of rationality within a complete information framework if
and only if it is optimal for a belief hierarchy within the corresponding
incomplete information framework that expresses common full belief in
caution, rationality, every good choice is supported, and prior belief in
the original utility functions.

This paper is organized as follows. Section \ref{sec:mac} gives a survey of
assumption of rationality in epistemic models with complete information and
the lexicographic epistemic models with incomplete information. Section \ref%
{sec:cce} gives the characterization result and their proofs. Section \ref%
{sec:cr} gives some concluding remarks on the relationship between the
result of this paper and characterization of permissbility in Section 4.6 of
Liu \cite{l17}.

\section{Models\label{sec:mac}}

\subsection{Complete information model}

In this subsection, we give a survey of lexicographic epistemic model with
complete information and define iterative admissibility within it. We adopt
the approach of Perea \cite{p12}, Chapters 5 and 7. See Brandenburger et al. 
\cite{bfk08} for an alternative approach. 

Consider a finite 2-person static game $\Gamma =(C_{i},u_{i})_{i\in I}$
where $I=\{1,2\}$ is the set of players, $C_{i}$ is the finite set of
choices and $u_{i}:C_{1}\times C_{2}\rightarrow \mathbb{R}$ is the utility
function for player $i\in I.$ In the following we sometimes denote $%
C_{1}\times C_{2}$ by $C$. We assume that each player has a lexicographic
belief on the opponent's choices, a lexicographic belief on the opponent's
lexicographic belief on her, and so on. This belief hierarchy is described
by a lexicographic epistemic model with types.\smallskip \newline
\textbf{Definition \ref{sec:mac}.1 (Epistemic model with complete
information).} Consider a finite 2-person static game $\Gamma
=(C_{i},u_{i})_{i\in I}$. A finite \emph{lexicographic epistemic model} for $%
\Gamma $ is a tuple $M^{co}=(T_{i},b_{i})_{i\in I}$ where\smallskip \newline
(a) $T_{i}$ is a finite set of types, and\smallskip \newline
(b) $b_{i}$ is a mapping that assigns to each $t_{i}\in T_{i}$ a
lexicographic belief over $\Delta (C_{j}\times T_{j}),$ i.e., $%
b_{i}(t_{i})=(b_{i1},b_{i2},...,b_{iK})$ where $b_{ik}\in \Delta
(C_{j}\times T_{j})$ for $k=1,...,K.$\smallskip 

Consider $t_{i}\in T_{i}$ with $b_{i}(t_{i})=(b_{i1},b_{i2},...,b_{iK}).$
Each $b_{ik}$ ($k=1,...,K$) is called $t_{i}$'s \emph{level-}$k$\emph{\
belief}. For each $(c_{j},t_{j})\in C_{j}\times T_{j},$ we say $t_{i}$ \emph{%
deems} $(c_{j},t_{j})$ \emph{possible} iff $b_{ik}(c_{j},t_{j})>0$ for some $%
k\in \{1,...,K\}.$ We say $t_{i}$ \emph{deems} $t_{j}\in T_{j}$ \emph{%
possible} iff $t_{i}$ deems $(c_{j},t_{j})$ possible for some $c_{j}\in
C_{j} $. For each $t_{i}\in T_{i},$ we denote by $T_{j}(t_{i})$ the set of
types in $T_{j}$ deemed possible by $t_{i}$. A type $t_{i}\in T_{i}$ is 
\emph{cautious} iff for each $c_{j}\in C_{j}$ and each $t_{j}\in
T_{j}(t_{i}),$ $t_{i}$ deems $(c_{j},t_{j})$ possible. That is, $t_{i}$
takes into account each choice of player $j$ for every belief hierarchy of $j
$ deemed possible by $t_{i}.$\smallskip

For each $c_{i}\in C_{i}$, let $%
u_{i}(c_{i},t_{i})=(u_{i}(c_{i},b_{i1}).,..,u_{i}(c_{i},b_{iK}))$ where for
each $k=1,...,K,$ $u_{i}(c_{i},b_{ik}):=\Sigma _{(c_{j},t_{j})\in
C_{j}\times T_{j}}b_{ik}(c_{j},t_{j})u_{i}(c_{i},c_{j}),$ that is, each $%
u_{i}(c_{i},b_{ik})$ is the expected utility for $c_{i}$ over $b_{ik}$ and $%
u_{i}(c_{i},t_{i})$ is a vector of expected utilities. For each $%
c_{i},c_{i}^{\prime }\in C_{i}$, we say that $t_{i}$ \emph{prefers} $c_{i}$ 
\emph{to} $c_{i}^{\prime }$, denoted by $u_{i}(c_{i},t_{i})>u_{i}(c_{i}^{%
\prime },t_{i}),$ iff there is $k\in \{0,...,K-1\}$ such that the following
two conditions are satisfied:\smallskip \newline
(a) $u_{i}(c_{i},b_{i\ell })=u_{i}(c_{i}^{\prime },b_{i\ell })$ for $\ell
=0,...,k,$ and\smallskip \newline
(b) $u_{i}(c_{i},b_{i,k+1})>u_{i}(c_{i}^{\prime },b_{i,k+1})$.\smallskip 
\newline
We say that $t_{i}$ \emph{is indifferent between }$c_{i}$ \emph{and }$%
c_{i}^{\prime },$ denoted by $u_{i}(c_{i},t_{i})=u_{i}(c_{i}^{\prime
},t_{i}),$ iff $u_{i}(c_{i},b_{ik})=u_{i}(c_{i}^{\prime },b_{ik})$ for each $%
k=1,...,K.$ It can be seen that the preference relation on $C_{i}$ under
each type $t_{i}$ is a linear order. $c_{i}$ is \emph{rational} (or \emph{%
optimal}) for $t_{i}$ iff $t_{i}$ does not prefer any choice to $c_{i}$.

For $(c_{j},t_{j}),(c_{j}^{\prime },t_{j}^{\prime })\in C_{j}\times T_{j},$
we say that $t_{i}$ \emph{deems} $(c_{j},t_{j})$ \emph{infinitely more
likely than} $(c_{j}^{\prime },t_{j}^{\prime })$ iff there is $k\in
\{0,...,K-1\}$ such that the following two conditions are
satisfied:\smallskip \newline
(a) $b_{i\ell }(c_{j},t_{j})=b_{i\ell }(c_{j}^{\prime },t_{j}^{\prime })=0$
for $\ell =1,...,k,$ and\smallskip \newline
(b) $b_{i,k+1}(c_{j},t_{j})>0$ and $b_{i,k+1}(c_{j}^{\prime },t_{j}^{\prime
})=0$.\smallskip \newline
\textbf{Definition \ref{sec:mac}.2 (Assumption of rationality) }A cautious
type\textbf{\ }$t_{i}\in T_{i}$ \emph{assumes the }$j$\emph{'s rationality}
iff the following two conditions are satisfied:\smallskip \newline
\textbf{(A1)} for all of player $j$'s choices $c_{j}$ that are optimal for
some cautious belief, $t_{i}$ deems possible some type $t_{j}$ for which $%
c_{j}$ is optimal;\smallskip \newline
\textbf{(A2)} $t_{i}$ deems all choice-type pairs $(c_{j},t_{j})$ where $%
t_{j}$ is cautious and $c_{j}$ is optimal for $t_{j}$ infinitely more likely
than any choice-type pairs $(c_{j}^{\prime },t_{j}^{\prime })$ that does not
have this property.\smallskip 

Informally speaking, assumption of the opponent's rationality is that $t_{i}$
puts all \textquotedblleft good\textquotedblright\ choices in front of those
\textquotedblleft bad\textquotedblright\ choices.The following definition
extends assumption of rationality inductively into $n$-fold for any $n\in 
\mathbb{N}.$\smallskip \newline
\textbf{Definition \ref{sec:mac}.3 (}$n$\textbf{-fold} \textbf{assumption of
rationality) }Consider a finite lexicographic epistemic model $%
M^{co}=(T_{i},b_{i})_{i\in I}$ for a game $\Gamma =(C_{i},u_{i})_{i\in I}$.
A cautious type $t_{i}\in T_{i}$ \emph{expresses }$1$\emph{-fold assumption
of rationality} iff it assumes $j$'s rationality. For any $n\in \mathbb{N},$
we say that a cautious type $t_{i}\in T_{i}$ expresses $(n+1)$\emph{-fold
assumption of rationality} iff the following two conditions are
satisfied:\smallskip \newline
\textbf{(nA1)} whenever a choice $c_{j}$ of player $j$ is optimal for some
cautious type (not necessarily in $M^{co}$) that expresses up to $n$-fold
assumption of rationality, $t_{i}$ deems possible some cautious type $t_{j}$
for player $j$ which expresses up to $n$-fold assumption of rationality for
which $c_{j}$ is optimal;\smallskip \newline
\textbf{(nA2)} $t_{i}$ deems all choice-type pair $(c_{j},t_{j})$, where $%
t_{j}$ is cautious and expresses up to $n$-fold assumption of rationality
and $c_{j}$ is optimal for $t_{j},$ infinitely more likely than any
choice-type pairs $(c_{j}^{\prime },t_{j}^{\prime })$ that does not satisfy
this property.\smallskip 

We say that $t_{i}$ \emph{expresses common assumption of rationality} iff it
expresses $n$-fold assumption of rationality for every $n\in \mathbb{N}.$

\subsection{Incomplete information model}

In this subsection, we give a survey of lexicographic epistemic model with
incomplete information defined in Liu \cite{l17} which is the counterpart of
the probabilistic epistemic model with incomplete information introduced by
Battigalli \cite{b03} and further developed in Battigalli and Siniscalchi 
\cite{bs03}, \cite{bs07}, and Dekel and Siniscalchi \cite{ds15}. We also
define some conditions on types in such a model.\smallskip \newline
\textbf{Definition \ref{sec:mac}.4 (Lexicographic epistemic model with
incomplete information)}. Consider a finite 2-person static game form $%
G=(C_{i})_{i\in I}.$ For each $i\in I,$ let $V_{i}$ be the set of utility
functions $v_{i}:C_{1}\times C_{2}\rightarrow \mathbb{R}.$ A \emph{finite
lexicographic epistemic model for }$G$\emph{\ with incomplete information}
is a tuple $M^{in}=(\Theta _{i},w_{i},\beta _{i})_{i\in I}$ where\smallskip 
\newline
(a) $\Theta _{i}$ is a finite set of types,\smallskip \newline
(b) $w_{i}$ is a mapping that assigns to each $\theta _{i}\in \Theta _{i}$ a
utility function $w_{i}(\theta _{i})\in V_{i},$ and\smallskip \newline
(c) $\beta _{i}$ is a mapping that assigns to each $\theta _{i}\in \Theta
_{i}$ a lexicographic belief over $\Delta (C_{j}\times \Theta _{j}),$ i.e., $%
\beta _{i}(\theta _{i})=(\beta _{i1},\beta _{i2},...,\beta _{iK})$ where $%
\beta _{ik}\in \Delta (C_{j}\times \Theta _{j})$ for $k=1,...,K.$\smallskip 

Concepts such as \textquotedblleft $\theta _{i}$ deems $(c_{j},\theta _{j})$
possible\textquotedblright\ and \textquotedblleft $\theta _{i}$ deems $%
(c_{j},\theta _{j})$ infinitely more likely than $(c_{j}^{\prime },\theta
_{j}^{\prime })$\textquotedblright\ can be defined in a similar way as in
Section \ref{sec:mac}.1. For each $\theta _{i}\in \Theta _{i},$ we use $%
\Theta _{j}(\theta _{i})$ to denote the set of types in $\Theta _{j}$ deemed
possible by $\theta _{i}$. For each $\theta _{i}\in \Theta _{i}$ and
\thinspace $v_{i}\in V_{i},$ $\theta _{i}^{v_{i}}$ is the auxiliary type
satisfying that $\beta _{i}(\theta _{i}^{v_{i}})=\beta _{i}(\theta _{i})$
and $w_{i}(\theta _{i}^{v_{i}})=v_{i}$.

For each $c_{i}\in C_{i},v_{i}\in V_{i},$ and $\theta _{i}\in \Theta _{i}$
with $\beta _{i}(\theta _{i})=(\beta _{i1},\beta _{i2},...,\beta _{iK}),$
let $v_{i}(c_{i},\theta _{i})=(v_{i}(c_{i},\beta
_{i1}),...,v_{i}(c_{i},\beta _{iK}))$ where for each $k=1,...,K,$ $%
v_{i}(c_{i},\beta _{ik}):=\Sigma _{(c_{j},\theta _{j})\in C_{j}\times \Theta
_{j}}\beta _{ik}(c_{j},\theta _{j})v_{i}(c_{i},c_{j}).$ For each $%
c_{i},c_{i}^{\prime }\in C_{i}$ and $\theta _{i}\in \Theta _{i},$ we say
that $\theta _{i}$ \emph{prefers }$c_{i}$ \emph{to} $c_{i}^{\prime }$ iff $%
w_{i}(\theta _{i})(c_{i},\theta _{i})>w_{i}(\theta _{i})(c_{i}^{\prime
},\theta _{i}).$ As in Section \ref{sec:mac}.1, this is also the
lexicographic comparison between two vectors. $c_{i}$ is \emph{rational} (or 
\emph{optimal}) for $\theta _{i}$ iff $\theta _{i}$ does not prefer any
choice to $c_{i}.$\smallskip \newline
\textbf{Definition \ref{sec:mac}.5 (Caution)}. $\theta _{i}\in \Theta _{i}$
is \emph{cautious} iff for each $c_{j}\in C_{j}$ and each $\theta _{j}\in
\Theta _{j}(\theta _{i})$, there is some utility function $v_{j}\in V_{j}$
such that $\theta _{i}$ deems $(c_{j},\theta _{j}^{v_{j}})$
possible.\smallskip 

This is a faithful translation of Perea and Roy \cite{ps17}'s definition of
caution in probabilistic model (p.312) into lexicographic model. It is the
counterpart of caution defined within the complete information framework in
Section \ref{sec:mac}.1; the only difference is that in incomplete
information models we allow different utility functions since $c_{j}$ will
be required to be rational for the paired type.\smallskip \newline
\textbf{Definition \ref{sec:mac}.6 (Belief in rationality)}. $\theta _{i}\in
\Theta _{i}$ \emph{believes in }$j$\emph{'s rationality }iff $\theta _{i}$
deems $(c_{j},\theta _{j})$ possible only if $c_{j}$ is rational for $\theta
_{j}.$\smallskip 

The following lemma shows that caution and a belief of full rationality can
be satisfied simultaneously in an incomplete information model because each
type is assigned with a belief on the opponent's choice-type pairs as well
as a payoff function. The consistency of caution and full rationality is the
essential difference between models with incomplete information and those
with complete information.\smallskip \newline
\textbf{Lemma \ref{sec:mac}.1 (Belief in rationality can be satisfied). }%
Consider a static game form $G=(C_{i})_{i\in I}$, $C_{i}^{\prime }\in C_{i}$%
, and $\beta _{i}=(\beta _{i1},\beta _{i2},...,\beta _{iK})$ such that $%
\beta _{ik}\in \Delta (C_{j})$ for each $k=1,...,K.$ Then there is $v_{i}\in
V_{i}$ such that each $c_{i}\in C_{i}^{\prime }$ is optimal in $v_{i}$ for $%
\beta _{i}.$\smallskip \newline
\textbf{Proof}. There are various way to construct such a $v_{i}.$ Here we
provide a simple one. For each $c\in C,$ let%
\begin{equation*}
v_{i}(c)=\left\{ 
\begin{array}{c}
1\text{ if }c_{i}\in C_{i}^{\prime }\text{ and }c_{j}\in \text{supp}\beta
_{i1}; \\ 
0\text{ \ \ \ \ \ \ \ \ \ \ \ \ \ \ \ \ \ \ \ \ \ \ \ \ \ \ \ otherwise}%
\end{array}%
\right. 
\end{equation*}%
It can be seen that each $c_{i}\in C_{i}^{\prime }$ is optimal in $v_{i}$
for $\beta _{i}.$ //\smallskip 

Caution and belief in rationality can be extended into $k$-fold for any $%
k\in \mathbb{N}$ as follows. Let $P$ be an arbitrary property of
lexicographic beliefs. We define that\smallskip \newline
(CP1) $\theta _{i}\in \Theta $ \emph{expresses }$0$\emph{-fold full belief in%
} $P$ iff $\theta _{i}$ satisfies $P;$\smallskip \newline
(CP2) For each $n\in \mathbb{N}$ with $n\geq 2,$ $\theta _{i}\in T_{i}$ 
\emph{expresses }$n$\emph{-fold full belief in} $P$ iff $\theta _{i}$ only
deems possible $j$'s types that express $n$-fold full belief in $P.$%
\smallskip 

$\theta _{i}$ \emph{expresses common full belief in} $P$ iff it expresses $n$%
-fold full belief in $P$ for each $n\in \mathbb{N}.$ By replacing $P$ with
\textquotedblleft caution\textquotedblright\ or \textquotedblleft
rationality\textquotedblright\ we can obtain common full belief in caution
or in rationality.\smallskip 

The following two conditions are important in characterizing assumption of
rationality.\smallskip \newline
\textbf{Definition \ref{sec:mac}.7 (Every good choice is supported)}.
Consider a static game form $G=(C_{i})_{i\in I},$ a\emph{\ }lexicographic
epistemic model $M^{in}=(\Theta _{i},w_{i},\beta _{i})_{i\in I}$ for\emph{\ }%
$G$\emph{\ }with incomplete information, and a pair $u=(u_{i})_{i\in I}$ of
utility functions. A cautious type $\theta _{i}\in \Theta _{i}$ \emph{%
believes in that every good choice of }$j$ \emph{is supported }iff for each $%
c_{j}$ that is optimal for some cautious type of $j$ (may not be in $M^{in}$%
) with $u_{j}$ as its assigned utility function, $\theta _{i}$ deems
possible a cautious type $\theta _{j}\in \Theta _{j}$ such that $%
w_{j}(\theta _{j})=u_{j}$ and $c_{j}$ is optimal for $\theta _{j}.$%
\smallskip \newline
\textbf{Definition \ref{sec:mac}.8 (Prior belief in }$u$\textbf{)}. Consider
a static game form $G=(C_{i})_{i\in I},$ a\emph{\ }lexicographic epistemic
model $M^{in}=(\Theta _{i},w_{i},\beta _{i})_{i\in I}$ for\emph{\ }$G$\emph{%
\ }with incomplete information, and a pair $u=(u_{i})_{i\in I}$ of utility
functions. $\theta _{i}\in \Theta _{i}$ \emph{priorly believes in }$u$ iff
for any $(c_{j},\theta _{j})$ with $\theta _{j}$ cautious deemed possible by 
$\theta _{i}$ satisfying that $w_{j}(\theta _{j})=u_{j},$ then $\theta _{i}$
deems $(c_{j},\theta _{j})$ infinitely more likely than any pair does not
satisfy that property.\smallskip 

Common full belief in that every good choice is supported and prior belief
in $u$ is different from that in caution or rationality. We have the
following definition.\smallskip \newline
\textbf{Definition \ref{sec:mac}.9 (}$n$\textbf{-fold} \textbf{belief in
that every good choice is supported and prior belief in }$u$\textbf{) }%
Consider a static game form $G=(C_{i})_{i\in I},$ a\emph{\ }lexicographic
epistemic model $M^{in}=(\Theta _{i},w_{i},\beta _{i})_{i\in I}$ for\emph{\ }%
$G$\emph{\ }with incomplete information, and a pair $u=(u_{i})_{i\in I}$ of
utility functions. $\theta _{i}\in \Theta _{i}$ \emph{express }$1$\emph{%
-fold belief in that every good choice is supported and prior belief in }$u$%
\emph{\ }iff it believes that every good choice of $j$ is supported and has
prior belief in $u$. For any $n\in \mathbb{N},$ we say that a cautious type $%
\theta _{i}\in \Theta _{i}$ expresses $(n+1)$\emph{-fold belief in prior
belief in that every good choice is supported and prior belief in }$u$ iff
the following two conditions are satisfied:\smallskip \newline
\textbf{(nP1)} whenever a choice $c_{j}$ of player $j$ is optimal for some
cautious type (not necessarily in $M^{in}$) with $u_{j}$ as its assigned
utility function that expresses up to $n$-fold belief in that every good
choice is supported, $\theta _{i}$ deems possible some cautious type $\theta
_{j}$ with $w_{j}(\theta _{j})=u_{j}$ for player $j$ which expresses up to $n
$-fold belief in that every good choice is supported and prior belief in $u$
for which $c_{j}$ is optimal.\smallskip \newline
\textbf{(nP2)} $\theta _{i}$ deems all choice-type pair $(c_{j},\theta _{j})$%
, where $\theta _{j}$ is cautious and expresses up to $n$-fold belief in
that prior belief in $u$ and every good choice is supported and satisfies $%
w_{j}(\theta _{j})=u_{j},$ infinitely more likely than any choice-type pairs 
$(c_{j}^{\prime },\theta _{j}^{\prime })$ that does not satisfy this
property.\smallskip 

We say that $t_{i}$ \emph{expresses common full belief in that every good
choice is supported and prior belief in }$u$ iff it expresses $n$-fold
belief in that every good choice is supported and prior belief in $u$ for
every $n\in \mathbb{N}.$

\section{Characterization\label{sec:cce}}

So far we have introduced two different groups of concepts for static games:
one includes assumption of rationality within a complete information
framework, the other contains some conditions on types within an incomplete
information framework. In this section we will show that there is
correspondence between them.\smallskip \newline
\textbf{Theorem \ref{sec:cce}.1 (Characterization of iterative admissibility)%
}. Consider a finite 2-person static game $\Gamma =(C_{i},u_{i})_{i\in I}$
and the corresponding game form $G=(C_{i}).$ $c_{i}^{\ast }\in C_{i}$ is
optimal to some type expressing common full belief in caution and common
assumption of rationality within some finite epistemic model with complete
information if and only if there is some finite epistemic model $%
M^{in}=(\Theta _{i},w_{i},\beta _{i})_{i\in I}$ with incomplete information
for $G$ and some $\theta _{i}^{\ast }\in \Theta _{i}$ with $w_{i}(\theta
_{i}^{\ast })=u_{i}$ such that\smallskip \newline
(a) $c_{i}^{\ast }$ is rational for $\theta _{i}^{\ast }$, and\smallskip 
\newline
(b) $\theta _{i}^{\ast }$ expresses common full belief in caution,
rationality, that every good choice is supported, and prior belief in $u$%
.\smallskip 

To show Theorem \ref{sec:cce}.1, we construct the mappings between finite
lexicographic epistemic models with complete information and those with
incomplete information. First, consider $\Gamma =(C_{i},u_{i})_{i\in I}$ and
a finite lexicographic epistemic model $M^{co}=(T_{i},b_{i})_{i\in I}$ with
complete information for $\Gamma .$ We first define types in a model with
incomplete information in the following two steps:\smallskip \newline
\textbf{Step 1}. For each $i\in I$ and $t_{i}\in T_{i},$ let $\Pi
_{i}(t_{i})=(C_{i1},...,C_{iL})$ be the partition of $C_{i}$ defined in
Lemma \ref{sec:mac}.1, that is, $\Pi _{i}(t_{i})$ is the sequence of
equivalence classes of choices in $C_{i}$ arranged from the most preferred
to the least preferred under $t_{i}.$ We define $v_{i\ell }(t_{i})\in V_{i}$
for each $\ell =1,...,L.$ We let $v_{i1}(t_{i})=u_{i}.$ By Lemma \ref%
{sec:mac}.1, for each $C_{i\ell }$ with $\ell >1$ there is some $v_{i\ell
}(t_{i})\in V_{i}$ such that each choice in $C_{i\ell }$ is rational at $%
v_{i\ell }(t_{i})$ under $t_{i}.$\smallskip \newline
\textbf{Step 2}. We define $\Theta _{i}(t_{i})=\{\theta
_{i1}(t_{i}),...,\theta _{iL}(t_{i})\}$ where for each $\ell =1,...,L,$ the
type $\theta _{i\ell }(t_{i})$ satisfies that (1) $w_{i}(\theta _{i\ell
}(t_{i}))=v_{i\ell }(t_{i}),$ and (2) $\beta _{i}(\theta _{i\ell }(t_{i}))$
is obtained from $b_{i}(t_{i})$ by replacing every $(c_{j},t_{j})$ with $%
c_{j}\in C_{jr}\in \Pi _{j}(t_{j})$ for some $r$ with $(c_{j},\theta _{j})$
where $\theta _{j}=\theta _{jr}(t_{j}),$ that is, $w_{j}(\theta _{j})$ is
the utility function among those corresponding to $\Pi _{j}(t_{j})$ in which 
$c_{j}$ is the rational for $t_{i}.$\smallskip 

For each $i\in I,$ let $\Theta _{i}=\cup _{t_{i}\in T_{i}}\Theta
_{i}(t_{i}). $ Here we have constructed a finite lexicographic epistemic
model $M^{in}=(\Theta _{i},w_{i},\beta _{i})_{i\in I}$ for the corresponding
game form $G=(C_{i})_{i\in I}$ with incomplete information. In the following
example we show how this construction goes.\smallskip \newline
\textbf{Example \ref{sec:cce}.2}. Consider the following game $\Gamma $
(Perea \cite{p12}, p.188):%
\begin{equation*}
\begin{tabular}{|l|l|l|}
\hline
$u_{1}\backslash u_{2}$ & $C$ & $D$ \\ \hline
$A$ & $1,0$ & $0,1$ \\ \hline
$B$ & $0,0$ & $0,1$ \\ \hline
\end{tabular}%
\end{equation*}%
and the lexicographic epistemic model $M^{co}=(T_{i},b_{i})_{i\in I}$ $%
\Gamma $ where $T_{1}=\{t_{1}\},$ $T_{2}=\{t_{2}\}$, and%
\begin{equation*}
b_{1}(t_{1})=((D,t_{2}),(C,t_{2})),\text{ }%
b_{2}(t_{2})=((A,t_{1}),(B,t_{1})).
\end{equation*}%
We show how to construct a corresponding model $M^{in}=(\Theta
_{i},w_{i},\beta _{i})_{i\in I}$. First, by Step 1 it can be seen that $\Pi
_{1}(t_{1})=(\{A\},\{B\})$ and $\Pi _{2}(t_{2})=(\{D\},\{C\}).$ We let $%
v_{11}(t_{1})=u_{1}$ where $A$ is rational for $t_{1}$ and $v_{12}(t_{1})$
where $B$ is rational for $t_{1}$ as follows. Similarly, we let $%
v_{21}(t_{2})=u_{2}$ where $D$ is rational under $t_{2}$ and $v_{22}(t_{2})$
where $C$ is rational under $t_{2}$ as follows:%
\begin{equation*}
\begin{tabular}{|l|l|l|}
\hline
$v_{12}(t_{1})$ & $C$ & $D$ \\ \hline
$A$ & $1$ & $0$ \\ \hline
$B$ & $0$ & $1$ \\ \hline
\end{tabular}%
,\text{ \ }%
\begin{tabular}{|l|l|l|}
\hline
$v_{22}(t_{2})$ & $C$ & $D$ \\ \hline
$A$ & $2$ & $1$ \\ \hline
$B$ & $0$ & $1$ \\ \hline
\end{tabular}%
.
\end{equation*}%
Then we go to Step 2. It can be seen that $\Theta _{1}(t_{1})=\{\theta
_{11}(t_{1}),\theta _{12}(t_{1})\},$ where%
\begin{eqnarray*}
w_{1}(\theta _{11}(t_{1})) &=&v_{11}(t_{1}),\text{ }\beta _{1}(\theta
_{11}(t_{1}))=((D,\theta _{21}(t_{2})),(C,\theta _{22}(t_{2}))), \\
w_{1}(\theta _{12}(t_{1})) &=&v_{12}(t_{1}),\text{ }\beta _{1}(\theta
_{12}(t_{1}))=((D,\theta _{21}(t_{2})),(C,\theta _{22}(t_{2}))).
\end{eqnarray*}%
Also, $\Theta _{2}(t_{2})=\{\theta _{21}(t_{2}),\theta _{22}(t_{2})\},$ where%
\begin{eqnarray*}
w_{2}(\theta _{21}(t_{2})) &=&v_{21}(t_{2}),\text{ }\beta _{2}(\theta
_{21}(t_{2}))=((A,\theta _{11}(t_{1})),(B,\theta _{12}(t_{1}))), \\
w_{2}(\theta _{22}(t_{2})) &=&v_{22}(t_{2}),\text{ }\beta _{2}(\theta
_{22}(t_{2}))=((A,\theta _{11}(t_{1})),(B,\theta _{12}(t_{1}))).
\end{eqnarray*}

Let $M^{co}=(T_{i},b_{i})_{i\in I}$ and $M^{in}=(\Theta _{i},w_{i},\beta
_{i})_{i\in I}$ be constructed from $M^{co}$ by the two steps above. We have
the following observations.\smallskip \newline
\textbf{Observation \ref{sec:cce}.1 (Redundancy)}. For each $t_{i}\in T_{i}$
and each $\theta _{i},\theta _{i}^{\prime }\in \Theta _{i}(t_{i}),$ $\beta
_{i}(\theta _{i})=\beta _{i}(\theta _{i}^{\prime }).$\smallskip \newline
\textbf{Observation \ref{sec:cce}.2 (Rationality)}. Eeach $\theta _{i}\in
\Theta _{i}(t_{i})$ believes in $j$'s rationality.\smallskip 

We omit their proofs since they hold by construction. Observation \ref%
{sec:cce}.1 means that the difference between any two types in a $\Theta
_{i}(t_{i})$ is in the utility functions assigned to them. Observation \ref%
{sec:cce}.2 means that in an incomplete information model constructed from
one with complete information, each type has (full) belief in the opponent's
rationality. This is because in the construction, we requires that for each
pair $(c_{j},t_{j})$ occurring in a belief, its counterpart in the
incomplete information replaces $t_{j}$ by the type in $\Theta _{j}(t_{j})$
with the utility function in which $c_{j}$ is optimal for $b_{i}(t_{j})$. It
follows from Observation \ref{sec:cce}.2 that each $\theta _{i}\in \Theta
_{i}(t_{i})$ expresses common full belief in rationality.\smallskip 

The following lemma shows that caution is preserved in this construction.%
\newline
\textbf{Lemma \ref{sec:cce}.1 (Caution}$^{co}\rightarrow $\textbf{\ Caution}$%
^{in}$\textbf{)}. Let $M^{co}=(T_{i},b_{i})_{i\in I}$ and $M^{in}=(\Theta
_{i},w_{i},\beta _{i})_{i\in I}$ be constructed from $M^{co}$ by the two
steps above. If $t_{i}\in T_{i}$ expresses common full belief in caution, so
does each $\theta _{i}\in \Theta _{i}(t_{i}).$\smallskip \newline
\textbf{Proof}. We show this statement by induction. First we show that if $%
t_{i}$ is cautious, then each $\theta _{i}\in \Theta _{i}(t_{i})$ is also
cautious. Let $c_{j}\in C_{j}$ and $\theta _{j}\in \Theta _{j}(\theta _{i}).$
By construction, it can be seen that the type $t_{j}\in T_{j}$ satisfying
the condition that $\theta _{j}\in \Theta _{j}(t_{j})$ is in $T_{j}(t_{i}).$
Since $t_{i}$ is cautious, $t_{i}$ deems $(c_{j},t_{j})$ possible. Consider
the pair $(c_{j},\theta _{j}^{\prime })$ in $\beta _{i}(\theta _{i})$
corresponding to $(c_{j},t_{j}).$ Since both $\theta _{j}$ and $\theta
_{j}^{\prime }$ are in $\Theta _{j}(t_{j}),$ it follows from Observation \ref%
{sec:cce}.1 that $\beta _{j}(\theta _{j})=\beta _{j}(\theta _{j}^{\prime }).$
Hence $(c_{j},\theta _{j}^{w_{j}(\theta _{j}^{\prime })})$ is deemed
possible by $\theta _{i}.$ Here we have shown that $\theta _{i}$ is cautious.

Suppose we have shown that, for each $i\in I,$ if $t_{i}$ expresses $n$-fold
full belief in caution then so does each $\theta _{i}\in \Theta _{i}(t_{i})$%
. Now suppose that $t_{i}$ expresses $(n+1)$-fold full belief in caution,
i.e., each $t_{j}\in T_{j}(t_{i})$ expresses $n$-fold full belief in
caution. By construction, for each $\theta _{i}\in \Theta _{i}(t_{i})$ and
each $\theta _{j}\in \Theta _{j}(\theta _{i})$ there is some $t_{j}\in
T_{j}(t_{i})$ such that $\theta _{j}\in \Theta _{j}(t_{i})$, and, by
inductive assumption, each $\theta _{j}\in \Theta _{j}(\theta _{i})$
expresses $n$-fold full belief in caution. Therefore, each $\theta _{i}\in
\Theta _{i}(t_{i})$ expresses $(n+1)$-fold full belief in caution.
//\smallskip 

We also need a mapping from epistemic models with incomplete information to
those with complete information. Consider a finite 2-person static game $%
\Gamma =(C_{i},u_{i})_{i\in I},$ the corresponding game form $%
G=(C_{i})_{i\in I},$ and a finite epistemic model $M^{in}=(\Theta
_{i},w_{i},\beta _{i})_{i\in I}$ for $G$ with incomplete information. We
construct a model $M^{co}=(T_{i},b_{i})_{i\in I}$ for $\Gamma $ with
complete information as follows. For each $\theta _{i}\in \Theta _{i},$ we
define $E_{i}(\theta _{i})=\{\theta _{i}^{\prime }\in \Theta _{i}:\beta
_{i}(\theta _{i}^{\prime })=\beta (\theta _{i})\}.$ In this way $\Theta _{i}$
is partitioned into some equivalence classes $\mathbb{E}_{i}=%
\{E_{i1},...,E_{iL}\}$ where for each $\ell =1,..,L,$ $E_{i\ell
}=E_{i}(\theta _{i})$ for some $\theta _{i}\in \Theta _{i}.$ To each $%
E_{i}\in \mathbb{E}_{i}$ we use $t_{i}(E_{i})$ to represent a type. We
define $b_{i}(t_{i}(E_{i}))$ to be a lexicographic belief which is obtained
from $\beta _{i}(\theta _{i})$ by replacing each occurrence of $%
(c_{j},\theta _{j})$ by $(c_{j},t_{j}(E_{j}(\theta _{j})));$ in other words, 
$b_{i}(t_{i}(E_{i}))$ has the same distribution on choices at each level as $%
\beta _{i}(\theta _{i})$ for each $\theta _{i}\in E_{i},$ while each $\theta
_{j}\in \Theta _{j}(\theta _{i})$ is replaced by $t_{j}(E_{j}(\theta _{j})).$
For each $i\in I,$ let $T_{i}=\{t_{i}(E_{i})\}_{E_{i}\in \mathbb{E}_{i}}.$
We have constructed from $M^{in}$ a finite epistemic model $%
M^{co}=(T_{i},b_{i})_{i\in I}$ with complete information for $\Gamma .$

It can be seen that this is the reversion of the previous construction. That
is, let $M^{co}=(T_{i},b_{i})_{i\in I}$ satisfying that $b_{i}(t_{i})\neq
b_{i}(t_{i}^{\prime })$ for each $t_{i},t_{i}^{\prime }\in T_{i}$ with $%
t_{i}\neq t_{i}^{\prime }$, and $M^{in}=(\Theta _{i},w_{i},\beta _{i})_{i\in
I}$ be constructed from $M^{co}$ by the previous two steps. Then $\mathbb{E}%
_{i}=\{\Theta _{i}(t_{i})\}_{t_{i}\in T_{i}}$ and $t_{i}(\Theta
_{i}(t_{i}))=t_{i}$ for each $i\in I.$

In the following example we show how this construction goes.\smallskip 
\newline
\textbf{Example \ref{sec:cce}.3}. Consider the game $\Gamma $ in Example \ref%
{sec:cce}.2 and the model $M^{in}=(\Theta _{i},w_{i},\beta _{i})_{i\in I}$
for the corresponding game form where $\Theta _{1}=\{\theta _{11},\theta
_{12}\},$ $\Theta _{2}=\{\theta _{21},\theta _{22}\},$ and%
\begin{eqnarray*}
w_{1}(\theta _{11}) &=&u_{1},\text{ }\beta _{1}(\theta _{11})=((D,\theta
_{21}),(C,\theta _{22})), \\
w_{1}(\theta _{12}) &=&v_{1},\text{ }\beta _{1}(\theta _{12})=((D,\theta
_{21}),(C,\theta _{22})), \\
w_{2}(\theta _{21}) &=&u_{2},\text{ }\beta _{2}(\theta _{21})=((A,\theta
_{11}),(B,\theta _{12})), \\
w_{2}(\theta _{22}) &=&v_{2},\text{ }\beta _{2}(\theta _{22})=((A,\theta
_{11}),(B,\theta _{12})).
\end{eqnarray*}%
where $v_{1}=v_{12}(t_{1})$ and $v_{2}=v_{22}(t_{2})$ in Example \ref%
{sec:cce}.2. It can be seen that $\mathbb{E}_{1}=\{\{\theta _{11},\theta
_{12}\}\}$ since $\beta _{1}(\theta _{11})=\beta _{1}(\theta _{12})$ and $%
\mathbb{E}_{2}=\{\{\theta _{21},\theta _{22}\}\}$ since $\beta _{2}(\theta
_{21})=\beta _{2}(\theta _{22}).$ Corresponding to those equivalence classes
we have $t_{1}(\{\theta _{11},\theta _{12}\})$ and $t_{2}(\{\theta
_{21},\theta _{22}\}),$ and

\begin{eqnarray*}
b_{1}(t_{1}(\{\theta _{11},\theta _{12}\})) &=&((D,t_{2}(\{\theta
_{21},\theta _{22}\})),(C,t_{2}(\{\theta _{21},\theta _{22}\}))), \\
b_{2}(t_{2}(\{\theta _{21},\theta _{22}\})) &=&((A,t_{1}(\{\theta
_{11},\theta _{12}\})),(B,t_{1}(\{\theta _{11},\theta _{12}\}))).
\end{eqnarray*}

We have the following lemmas.\smallskip \newline
\textbf{Lemma \textbf{\ref{sec:cce}}.2} \textbf{(Caution}$^{in}\rightarrow $%
\textbf{\ Caution}$^{co}$\textbf{)}. Let $M^{in}=(\Theta _{i},w_{i},\beta
_{i})_{i\in I}$ and $M^{co}=(T_{i},b_{i})_{i\in I}$ be constructed from $%
M^{in}$ by the above approach. If $\theta _{i}\in \Theta _{i}$ expresses
common full belief in caution, so does $t_{i}(E_{i}(\theta _{i})).$%
\smallskip \newline
\textbf{Proof}. We show this statement by induction. First we show that if $%
\theta _{i}$ is cautious, then $t_{i}(E_{i}(\theta _{i}))$ is also cautious.
Let $c_{j}\in C_{j}$ and $t_{j}\in T_{j}(t_{i}(E_{i}(\theta _{i}))).$ By
construction, $t_{j}=t_{j}(E_{j})$ for some $E_{j}\in \mathbb{E}_{j},$ and
there is some $\theta _{j}\in E_{j}$ which is deemed possible by $\theta
_{i}.$ Since $\theta _{i}$ is cautious, there is some $\theta _{j}^{\prime }$
with $\beta _{j}(\theta _{j}^{\prime })=\beta _{j}(\theta _{j}),$ i.e., $%
\theta _{j}^{\prime }\in E_{j},$ such that $(c_{j},\theta _{j}^{\prime })$
is deemed possible by $\theta _{i}.$ By construction, $(c_{j},t_{j})$ is
deemed possible by $t_{i}(E_{i}(\theta _{i})).$

Suppose we have shown that, for each $i\in I,$ if $\theta _{i}$ expresses $n$%
-fold full belief in caution then so does $t_{i}(E_{i}(\theta _{i}))$. Now
suppose that $\theta _{i}$ expresses $(n+1)$-fold full belief in caution,
i.e., each $\theta _{j}\in \Theta _{j}(\theta _{i})$ expresses $n$-fold full
belief in caution. Since, by construction, for each $t_{j}\in
T_{j}(t_{i}(E_{i}(\theta _{i})))$, there is some $\theta _{j}\in \Theta
_{j}(\theta _{i})$ such that $t_{j}=t_{j}(E_{j}(\theta _{j})),$ by inductive
assumption $t_{j}$ expresses $n$-fold full belief in caution. Therefore, $%
t_{i}(E_{i}(\theta _{i}))$ expresses $(n+1)$-fold full belief in caution.
//\smallskip \newline
\textbf{Lemma \ref{sec:cce}.3 (Assumption of rationality }$%
\longleftrightarrow $\textbf{\ every good choice is supportedprior\ + belief
in }$u$ \textbf{)}. Let $M^{co}=(T_{i},b_{i})_{i\in I}$ and $M^{in}=(\Theta
_{i},w_{i},\beta _{i})_{i\in I}$ be constructed from $M^{co}$. If $t_{i}\in
T_{i}$ expresses common assumption of rationality, then each $\theta _{i}\in
\Theta _{i}(t_{i})$ expresses common full belief in that every good choice
is supported and prior belief in $u$.

On the other hand, let $M^{in}=(\Theta _{i},w_{i},\beta _{i})_{i\in I}$ and $%
M^{co}=(T_{i},b_{i})_{i\in I}$ be constructed from $M^{in}$. If $\theta
_{i}\in \Theta _{i}$ expresses common full belief in that every good choice
is supported and prior belief in $u$, then $t_{i}(E_{i}(\theta _{i}))$
expresses common full assumption of rationality.\smallskip \newline
\textbf{Proof}. We show this statement by induction. Let $\theta _{i}\in
\Theta _{i}(t_{i}).$ First we show that if $t_{i}$ assumes in $j$'s
rationality, $\theta _{i}$ believes that every good choice is supported and
prior belief in $u$. Let $c_{j}\in C_{j}$ be optimal for some cautious type
of $j$ whose assigned utility function is $u_{j}$ within an epistemic model
with incomplete information. It is easy to see that $c_{j}$ is optimal for
its corresponding type, which is also cautious by Lemma \ref{sec:cce}.2, in
any complete information model constructed from the one with incomplete
information by our approach above. Since $t_{i}$ assumes $j$'s rationality, $%
t_{i}$ deems possible a cautious type $t_{j}$ for which $c_{j}$ is optimal.
By construction, some $\theta _{j}\in \Theta _{j}(t_{j})$ is deemed possible
by $\theta _{i}.$ Since $t_{i}$ is cautious, $(c_{j},t_{j})$ is deemed
possible by $t_{i},$ and, by construction $(c_{j},\theta _{j1}(t_{j}))$ is
deemed possible by $\theta _{i}$. Since $w_{j}(\theta _{j1}(t_{j}))=u_{j}$
and $c_{j}$ is optimal for $\theta _{j1}(t_{j}),$ it follows that $\theta
_{i}$ believes in that every good choice is supported.

Let $(c_{j},\theta _{j})$ with $\theta _{j}$ cautious deemed possible by $%
\theta _{i}$ satisfying $w_{j}(\theta _{j})=u_{j}$ and $(c_{j}^{\prime
},\theta _{j}^{\prime })$ a pair which does not satisfy that condition. Let $%
(c_{j},t_{j})$ and $(c_{j}^{\prime },t_{j}^{\prime })$ be the pairs occuring
in the belief of $t_{i}$ corresponding to $(c_{j},\theta _{j})$ and $%
(c_{j}^{\prime },\theta _{j}^{\prime }).$ Since $c_{j}$ is rational to $%
\theta _{j}$ and $w_{j}(\theta _{j})=u_{j},$ it follows that $c_{j}$ is
optimal for $t_{j}.$ On the other hand, $c_{j}^{\prime }$ is not optimal for 
$t_{j}^{\prime }.$ Since $t_{i}$ assumes $j$'s rationality, $t_{i}$ deems $%
(c_{j},t_{j})$ infinitely more likely than $(c_{j}^{\prime },t_{j}^{\prime
}).$ By construction, $\theta _{i}$ deems $(c_{j},\theta _{j})$ infinitely
more likely than $(c_{j}^{\prime },\theta _{j}^{\prime }).$ Here we have
shown that $\theta _{i}$ priorly believes in $u.$\smallskip 

Now we show the other direction: suppose that if $\theta _{i}\in \Theta _{i}$
believes in that every good choice is supported and priorly believes in $u$,
we prove that $t_{i}(E_{i}(\theta _{i}))$ assumes $j$'s rationality. Suppose
that $c_{j}$ is optimal for some cautious type within some epistemic model
with complete information. It can be seen by construction that $c_{j}$ is
optimal for some cautious type with $u_{i}$ as its assigned utility function
within some epistemic model with incomplete information which corresponds to
that complete information model. Since $\theta _{i}$ believes in that every
good choice is supported, $\theta _{i}$ deems possible a cautious type $%
\theta _{j}$ such that $w_{j}(\theta _{j})=u_{j}$ and $c_{j}$ is optimal for 
$\theta _{j}.$ By construction it follows that $t_{i}(E_{i}(\theta _{i}))$
deems $t_{j}(E_{j}(\theta _{j}))$ possible for which $c_{j}$ is optimal.

Let $(c_{j},t_{j})$ with $t_{j}$ cautious be a pair which is deemed possible
by $t_{i}(E_{i}(\theta _{i}))$ satisfying that $c_{j}$ is optimal for $t_{j},
$ and $(c_{j}^{\prime },t_{j}^{\prime })$ be a pair deemed possible by $%
t_{i}(E_{i}(\theta _{i}))$ which does not satisfy that condition. Let $%
(c_{j},\theta _{j})$ and $(c_{j}^{\prime },\theta _{j}^{\prime })$ be the
corresponding pairs occuring in the belief of $\theta _{i}.$ Since $\theta
_{i}$ believes in rationality, by construction it follows that $u_{j}(\theta
_{j})=u_{j}$ while $u_{j}(\theta _{j}^{\prime })\neq u_{j}.$ Since $\theta
_{i}$ priorly believes in $u,$ $\theta _{i}$ deems $(c_{j},\theta _{j})$
infinitely more likely than $(c_{j}^{\prime },\theta _{j}^{\prime })$. It
follows that $t_{i}(E_{i}(\theta _{i}))$ deems $(c_{j},t_{j})$ infinitely
more likely than $(c_{j}^{\prime },t_{j}^{\prime }).$ Here we have shown
that $t_{i}(E_{i}(\theta _{i}))$ assumes $j$'s rationality.\smallskip 

Suppose that, for some $n\in \mathbb{N},$ we have shown that for each $k\leq
n,$\smallskip \newline
(n1) if $t_{i}\in T_{i}$ expresses $k$-fold assumption of rationality, then
each $\theta _{i}\in \Theta _{i}(t_{i})$ expresses $k$-fold full belief in
that every good choice is supported and prior belief in $u$;\smallskip 
\newline
(n2) If $\theta _{i}\in \Theta _{i}$ expresses $k$-fold full belief in that
every good choice is supported and prior belief in $u$, then $%
t_{i}(E_{i}(\theta _{i}))$ expresses $k$-fold assumption of
rationality.\smallskip 

Now we show that these two statements hold for $n+1.$ First, suppose that $%
t_{i}\in T_{i}$ expresses $(n+1)$-fold assumption of rationality. Let $%
c_{j}\in C_{j}$ be a choice of $j$ optimal for some cautious type whose
assigned utility function is $u_{j}$ that expresses up to $n$-fold belief in
that every good choice is supported. Then it is easy to see that (1) by
inductive assumption, in the constructed complete information model the
corresponding type expresses $n$-fold assumption of rationality, and (2) $%
c_{j}$ is optimal for that type. Since $t_{i}$ expresses $(n+1)$-fold
assumption of rationality, $t_{i}$ deems possible a cautious type $t_{j}$
that expresses up to $n$-fold assumption of rationality and for which $c_{j}$
is optimal. By construction, it follows that $\theta _{i}$ deems possible
some $\theta _{j}\in \Theta _{j}(t_{j})$. By inductive assumption it follows
that each $\theta _{j}\in \Theta _{j}(t_{j})$ expresses $n$-fold belief in
that every good choice is supported. Since $\theta _{i}$ expresses common
belief in caution and rationality it follows that $\theta _{i}$ deems $%
(c_{j},\theta _{j1})$ for $\theta _{j1}\in \Theta _{j}(t_{j})$ (that is, $%
w_{j}(\theta _{j1})=u_{j}$).

Let $(c_{j},\theta _{j})$ with $\theta _{j}$ cautious deemed possible by $%
\theta _{i}$ satisfying that $\theta _{j}$ expresses up tp $n$-fold belief
in prior belief in $u$ and that every good choice is supported and $%
w_{j}(\theta _{j})=u_{j}$ and $(c_{j}^{\prime },\theta _{j}^{\prime })$ a
pair which does not satisfy those conditions. Let $(c_{j},t_{j})$ and $%
(c_{j}^{\prime },t_{j}^{\prime })$ be the pairs occuring in the belief of $%
t_{i}$ corresponding to $(c_{j},\theta _{j})$ and $(c_{j}^{\prime },\theta
_{j}^{\prime }).$ Since $c_{j}$ is rational for $\theta _{j}$ and $%
w_{j}(\theta _{j})=u_{j},$ it follows that $c_{j}$ is optimal for $t_{j}.$
Also, by inductive assumption, it follows that $t_{j}$ expresses up to $n$%
-fold assumption of rationality. On the other hand, it can be seen that $%
(c_{j}^{\prime },t_{j}^{\prime })$ does not satisfy these conditions. Since $%
t_{i}$ expresses $(n+1)$-fold of assumptions of rationality, $t_{i}$ deems $%
(c_{j},t_{j})$ infinitely more likely than $(c_{j}^{\prime },t_{j}^{\prime
}).$ By construction, $\theta _{i}$ deems $(c_{j},\theta _{j})$ infinitely
more likely than $(c_{j}^{\prime },\theta _{j}^{\prime }).$ Here we have
shown that $\theta _{i}$ expresses $(n+1)$-fold full belief in that every
good choice is supported and prior belief in $u$.\smallskip 

Now suppose that $\theta _{i}\in \Theta _{i}$ expresses $(n+1)$-fold full
belief in that every good choice is supported and prior belief in $u$. Let $%
c_{j}\in C_{j}$ be a choice of $j$ optimal for some cautious type that
expresses to $n$-fold assumption of rationality. By inductive assumption it
follows that the corresponding type within some incomplete information model
also expresses $n$-fold full belief in that every good choice is supported
and prior belief in $u$. It can be seen that $c_{j}$ is optimal to the
constructed type having $u_{j}$ as its utility functionand the type
expresses up to $n$-fold full belief in that every good choice is supported
and prior belief in $u$. Then $\theta _{i}$ deems possible a type $\theta
_{j}$ with $w_{j}(\theta _{j})=u_{j}$ for player $j$ which expresses up to $n
$-fold belief in that every good choice is supported for which $c_{j}$ is
optimal. By inductive assumption it follows that $t_{i}(E_{i}(\theta _{i}))$
deems possible $t_{j}(E_{j}(\theta _{j}))$ which expresses $n$-fold
assumption of rationality and for which $c_{j}$ is optimal.

Let $(c_{j},t_{j}$) be a pair with $t_{j}$ cautious deemed possible by $%
t_{i}(E_{i}(\theta _{i}))$ where $t_{j}$ expresses up to $n$-fold assumption
of rationality and $c_{j}$ is optimal for $t_{j}$, and let $(c_{j}^{\prime
},t_{j}^{\prime })$ be a pair that does not satisfy this property. Let $%
(c_{j},\theta _{j})$ and $(c_{j}^{\prime },\theta _{j}^{\prime })$ be the
corresponding pairs occurring in the belief of $\theta _{i}.$ By inductive
assumption and by construction, $\theta _{j}$ is cautious and expresses up
to $n$-fold belief in that prior belief in $u$ and every good choice is
supported and $w_{j}(\theta _{j})=u_{j},$ while $(c_{j}^{\prime },\theta
_{j}^{\prime })$ does not satisfy this property. Therefore $\theta _{i}$
deems $(c_{j},\theta _{j})$ infinitely more likely than $(c_{j}^{\prime
},\theta _{j}^{\prime }),$ which implies that $t_{i}(E_{i}(\theta _{i}))$
deems $(c_{j},t_{j})$ infinitely more likely than $(c_{j}^{\prime
},t_{j}^{\prime }).$ Here we have shown that $t_{i}(E_{i}(\theta _{i}))$
expresses $(n+1)$-fold assumption of rationality. //\smallskip \newline
\textbf{Proof of Theorem \ref{sec:cce}.1}. \textbf{(Only-if) }Let $%
M^{co}=(T_{i},b_{i})_{i\in I}$, $M^{in}=(\Theta _{i},w_{i},\beta _{i})_{i\in
I}$ be constructed from $M^{co}$ by the two steps above, $c_{i}^{\ast }\in
C_{i}$ be a permissible choice, and $t_{i}^{\ast }\in T_{i}$ be a type
expressing common full belief in caution and common assumption of
rationality such that $c_{i}^{\ast }$ is rational for $t_{i}^{\ast }.$ Let $%
\theta _{i}^{\ast }=\theta _{i1}(t_{i}^{\ast }).$ By definition, $%
w_{i}(\theta _{i}^{\ast })=u_{i}$ and $\beta _{i}(\theta _{i}^{\ast })$ has
the same distribution on $j$'s choices at each level as $b_{i}(t_{i}^{\ast })
$. Hence $c_{i}^{\ast }$ is rational for $\theta _{i}^{\ast }.$ Also, it
follows from Observation \ref{sec:cce}.2, Lemmas \ref{sec:cce}.1, and \ref%
{sec:cce}.3 that $\theta _{i}^{\ast }$ expresses common full belief in
caution, rationality,that a good choice is supported, and prior belief in $u$%
.

\textbf{(If)}. Let $M^{in}=(\Theta _{i},w_{i},\beta _{i})_{i\in I}$, $%
M^{co}=(T_{i},b_{i})_{i\in I}$ be constructed from $M^{in}$ by the above
approach, and $c_{i}^{\ast }\in C_{i}$ be rational for some $\theta
_{i}^{\ast }$ with $w_{i}(\theta _{i}^{\ast })=u_{i}$ which expresses common
full belief in caution, rationality, that a good choice is supported, and
prior belief in $u$. Consider $t_{i}(E_{i}(\theta _{i}^{\ast })).$ Since $%
w_{i}(\theta _{i}^{\ast })=u_{i}$ and $b_{i}(t_{i}(E_{i}(\theta _{i}^{\ast
})))$ has the same distribution on $j$'s choices at each level as $\beta
_{i}(\theta _{i}^{\ast })$, $c_{i}^{\ast }$ is rational for $%
t_{i}(E_{i}(\theta _{i}^{\ast })).$ Also, by Lemmas \ref{sec:cce}.2 and \ref%
{sec:cce}.3, $t_{i}(E_{i}(\theta _{i}^{\ast }))$ expresses common full
belief in caution and common assumption of rationality. //

\section{Concluding Remarks\label{sec:cr}}

Assumption of rationality is a refinement of permissibility (See Perea \cite%
{p12}). This can also be seen within the framework of incomplete
information. Comparing our characterization of the former of the
characterization of the latter in Section 4.6 in Liu \cite{l17} it can be
seen that there is correspondence between the conditions. Section 4.6 in Liu 
\cite{l17} characterizes permissibility by weak caution, rationality, and
primary belief in $u$ within the incomplete information framework. The
characteization of assumption of rationality shares rationality with it,
while caution and prior belief are stronger than weak caution and primary
belief in $u,$ respectively.

An interesting phenomenon is the role of rationality. Liu \cite{l17}
provides two ways to characterize permissibility, one with rationality and
one without it. The characterization of proper rationality there is a
stronger version of the latter, while the characterization in this paper a
stronger version of the former. So far, it seems that using or not using
rationality in the characterization differentiate the two refinements of
permissibility, that is, assumption of rationality and proper
rationalizability, within the incomplete information framework. It would be
interesting that any future research would confirm this statement or provide
any counterexample, that is, show that proper rationalizability can be
characterized with rationality while assumption of rationality can be done
without it.

On the other hand, as shown in Liu \cite{l17} (and the construction here),
it is always possible to construct epistemic models with incomplete
information which satisfies rationality as well as all conditions for
characterization of proper rationalizability.  Further, prior belief in $u$
is voguely a condition between primary belief in $u$ and $u$-centered belief
which is used in Theorem 3.2 of Liu \cite{l17} to characterize proper
rationalizability. Those seem to correspond to the fact within the complete
information framework that there is always possible to construct belief
hierarchy which both assumes the opponent's rationality and respects the
opponent's preferences.

\section*{Acknowledgement}

The author would like to thank Andr\'{e}s Perea for his valuable discussion
and encouragement. She thanks all teachers and students in the 4th Epicenter
Spring Course on Epistemic Game Theory at Maastricht University to whom she
owes inspiring teaching, stimulating discussions, and new ideas. She
gratefully acknowledges the support of Grant-in-Aids for Young Scientists
(B) of JSPS No.17K13707 and Grant for Special Research Project No. 2017K-016
of Waseda University.

{}

\end{document}